\documentclass[12pt]{article}
\usepackage{amssymb,amscd,amsmath}

\newcommand{\bm}{j_{-}}
\newcommand{\bp}{j_{+}}
\newcommand{\ap}{\alpha_{+}}
\newcommand{\am}{\alpha_{-}}
\newcommand{\bz}{\bar z}
\newcommand{\bx}{\bar x}
\newcommand{\ip}{i^{\prime}}

\newcommand{\ls}{\hat {sl}_2}
\newcommand{\po}{\Phi^{j\,\bar j}}
\newcommand{\pf}{\Phi^{j}}
\newcommand{\prh}{\rho^{\prime}}
\def\mo#1#2{\Phi_{#1.#2}}
\def\fzx#1{\Phi^{j_{#1}}(x_{#1},\bx_{#1},z_{#1},\bz_{#1})}
\def\dj#1{\Delta_{j_{#1}}}
\def\ff#1{\Phi^{j_{#1}}}

\def\h{\frac{1}{2}}
\def\sp{\sigma^{\prime}}
\def\j#1#2{j_{#1.#2}}
\def\C{\mathbb{C}}

\def\N{\mathbb{N}}
\def\R{\mathbb{R}}
\def\NP#1#2{{\it Nucl.Phys.} {\bf B#1} (#2)}
\def\PL#1#2{{\it Phys.Lett.} {\bf B#1} (#2)}
\def\CMP#1#2{{\it Commun.Math.Phys.} {\bf #1} (#2)}
\title{ Operator Algebra of the SL(2) conformal field theories}
\author{Oleg Andreev\thanks{e-mail: andre@landau.ac.ru}
\thanks{This work was supported in part by Russian Basic Research Foundation
under grant 93-02-3135} \\ \\
L.D.Landau Institute for Theoretical Physics,\\
Kosygina 2, 117940 Moscow, Russia}
\date{}
\begin{document}
\maketitle

\begin{abstract}
Structure constants of Operator Algebras for the SL(2) degenerate conformal
field theories are calculated.
\end{abstract}

\vspace{-11cm}
\hfill LANDAU-95-TMP-1

\hfill hep-th/9504082

\vspace{11.5cm}
Since the seminal work of Belavin, Polyakov and Zamolodchikov \cite{BPZ},
there has been much progress in understanding two-dimensional
conformal field theories. It is essential to compute the structure
constants of Operator Algebra of such theories. In fact, it was done only for
relatively few theories. The most famous examples are the diagonal minimal
models and SU(2) WZW theory \cite{DF,FZ}. There are also works on the structure
constants of non diagonal theories (see e.g.\cite{CF} and refs. therein).

Degenerate conformal theories attract special interest since they are known to
describe critical fluctuations in statistical models. From a mathematical point
of view they correspond to reducible (with singular vectors) representations of
chiral algebras. An irreducible representations are obtained by setting the
singular vectors to zero that leads to differential equations for correlation
functions \cite{BPZ}.

I this paper, I shall compute the structure constants of Operator Algebra for
the SL(2) degenerate conformal field theories. Such theories contain
the reducible representations of the chiral algebra $\ls$ with the
highest weights listed by Kac-Kazhdan (see (5) below).
The well-known integrable representations are a special case, namely
$j^+_{1.m}\,k\in\N$.
These theories are of great interest because of their relevance in 2d quantum
gravity coupled to conformal minimal matter as well as their connection via
a quantum hamiltonian reduction with Dotsenko-Fateev (DF) models.

The theories have $\ls\times\ls$
algebra as the symmetry algebra. The chiral currents of $\ls$ form the
following
OP algebra
\begin{align}
J^{\alpha}(z_1)J^{\beta}(z_2)=\frac{k}{2}q^{\alpha\beta}\frac{1}{(z_1-z_2)^2}+
\frac{f^{\alpha\beta}_{\gamma}}{(z_1-z_2)}J^{\gamma}(z_2)+O(1)\qquad,
\end{align}
where $k\text{ is the
level},\,\,q^{00}=1,\,\,q^{+-}=q^{-+}=2,\,\,f^{0+}_{+}=f^{-0}_{-}=1,\,\,
f^{+-}_{0}=2;\,\,\alpha,\beta=0,+,-$.
\newline The same OP expansions, of course, are valid for antiholomorphic
currents.

The stress-energy tensor of the models has two independent components which can
be chosen in the Sugawara form
\begin{align}
\begin{split}
T(z)=\frac{1}{k+2}q_{\alpha\beta}:J^{\alpha}(z)J^{\beta}(z):\qquad,\\
\bar T(\bz)=\frac{1}{k+2}q_{\alpha\beta}:\bar
J^{\alpha}(\bz)\bar J^{\beta}(\bz):\qquad.
\end{split}
\end{align}
The primary fields (basic conformal operators) are defined as
\begin{align}
\begin{split}
J^{\alpha}(z_1)\po(z_2,\bz_2)=\frac{S_j^{\alpha}}{z_1-z_2}\po(z_2,\bz_2)+O(1)
\qquad,\\
\bar J^{\alpha}(\bz_1)\po(z_2,\bz_2)=\frac{\bar S_{\bar
j}^{\alpha}}{\bz_1-\bz_2}
\po(z_2,\bz_2)+O(1)\qquad.
\end{split}
\end{align}
Here $S_j^{\alpha}(\bar S_{\bar j}^{\alpha})$ is the "left"("right")
representation of $sl_2$. The conformal dimensions of $\po$ follow from the OP
expansions with $T(z),\,\bar T(\bz)$. They are given by
$\Delta_j=\frac{j(j+1)}{k+2}$ and $\bar\Delta_{\bar j}=\frac{\bar j(\bar
j+1)}{k+2}$, respectively.

The complete system of local fields involved in the theory includes, besides
the primary fields $\po$, all the fields of the form
\begin{align}
J^{\alpha_1}_{n_1}\dots J^{\alpha_N}_{n_N}
\bar J^{\beta_1}_{\bar n_1}\dots\bar J^{\beta_M}_{\bar n_M}\po\qquad,
\end{align}
where $J^{\alpha}_{n},\,\,\bar J^{\beta}_{\bar n}$ are the Laurent series
components of $J^{\alpha}(z)\text{ and }\bar J^{\beta}(\bz)$, respectively.
Following \cite{BPZ}, I shall denote by $[\po]$ the hole set of the fields (4)
associated with the primary field $\po$. From mathematical point of view
$[\po]$ corresponds to the highest weight representation of $\ls\times\ls$.

In this work I will consider only the diagonal embedding the physical space of
states into a tensor product of two chiral space of states. Such models are
known as "A" series. Since for these models all primary fields are spinless,
i.e. $\bar j\equiv j(\bar \Delta\equiv\Delta)$, I suppress $\bar j$-dependence
below.

In \cite{KK} Kac and Kazhdan found that the highest weight representation
of $\ls$ is reducible if the highest weight $j$ takes the values $\j{n}{m}$
defined by
\begin{align}
j^+_{n.m}=\frac{n-1}{2}\bm+\frac{m-1}{2}\bp\,\qquad\text{ or }\qquad
j^-_{n.m}=-\frac{n}{2}\bm-\frac{m+1}{2}\bp\quad,
\end{align}
with $\bp=1\,,\,\bm=-k-2\,,\,k\in\C\,,\,\{n,m\}\in\N$.

In general, given representations of a chiral algebra(symmetry algebra), to
define fields of conformal field theory, one needs a construction attaching
representation to a point. In \cite{FM} Feigin and Malikov proposed the
improved
construction for the weights given by (5). The point is that one more parameter
should be introduced and a module should be attached to a pair. The first
parameter is a point on a curve. As to the second, it can be taken as
an isotopic coordinate \footnote{It should be noted that the primary fields
including dependence on the isotopic coordinate have also been considered
by Furlan et al.\cite{PG}.}. Actually, this has a
very simple physical interpretation.  Since $\Delta$ are quadratic in $j$ one
has to introduce additional parameter in order to define OP expansions (12)
unambiguously otherwise they are defined up to $j=-j-1$ identification.
Note that for the integrable representations the OP
algebra is closed for $0\leq j\leq k/2$ with $k\in\N$ so the OP expansions are
well defined without $x$. However this is not the case for a general $j$
defined by (5).

Let me now define $S_j^{\alpha}$ as
\begin{align}
S_j^{-}=\frac{\partial}{\partial x}\,,\qquad S_j^{0}=
-x\frac{\partial}{\partial x}+j\,,\qquad
S_j^{+}=-x^2\frac{\partial}{\partial x}+2jx\,,
\end{align}
The same definitions with substitutions $x\rightarrow\bx,\,S_j^{\alpha}
\rightarrow\bar S_j^{\alpha}$ are also valid.
In above the isotopic coordinates $x,\bx$ were introduced. Together with
$z,\bz$ they form the Malikov-Feigin pair.

The chiral currents (1) can be turned into a form(current)
\begin{align}
J(x,z)=J^+(z)-2xJ^0(z)-x^2J^-(z)\qquad.
\end{align}
It is easily shown that the OP expansion of $J(x,z)$ is
\begin{align}
J(x_1,z_1)J(x_2,z_2)=-k\frac{x_{12}^2}{z_{12}^2}-2\frac{x_{12}}{z_{12}}
J(x_2,z_2)-\frac{x_{12}^2}{z_{12}}\frac{\partial}{\partial x_2}J(x_2,z_2)+O(1)
\qquad,
\end{align}
where $z_{ij}=z_i-z_j$. The same OP expansion, of course, is valid for
antiholomorphic current.

Define the primary fields as
\begin{align}
J(x_1,z_1)\pf(x_2,\bx_2,z_2,\bz_2)=-2j\frac{x_{12}}{z_{12}}
\pf(x_2,\bx_2,z_2,\bz_2)-\frac{x_{12}^2}{z_{12}}\frac{\partial}{\partial x_2}
\pf(x_2,\bx_2,z_2,\bz_2)+O(1)\quad.
\end{align}
\newline It should be noted that in general case the primary fields are
non-polynomial in $x,\bx$. Furthermore, $J(x,z)$ is not primary.

The highest weight representations of $\ls\times\ls$ are built as in
(4) with substitutions $J^{\alpha}_n\rightarrow J^{\alpha}_n(x),\, \bar
J^{\alpha}_{\bar n}\rightarrow\bar J^{\alpha}_{\bar n}(\bx)\text{ and
}\po\rightarrow\pf$. The $x(\bx)$-dependent components of $J(x,z)(\bar
J(\bx,\bz))$ are given by
\begin{align}
J^-_n(x)=J^-_n\quad ,\quad J^0_n(x)=J^0_n+xJ^-_n\quad ,\quad
J^+_n(x)=J^+_n-2xJ^0_n-x^2J_n^-\quad.
\end{align}
It is evident that $J^{\alpha}(x)$ form the Kac-Moody algebra
\begin{align*}
[J^{\alpha}_n(x),J^{\beta}_m(x)]=f^{\alpha\beta}_{\gamma}J^{\gamma}_{n+m}(x)
+\frac{k}{2}nq^{\alpha\beta}\delta_{n+m}\quad.
\end{align*}

The Operator Product of any two operators is given by
\begin{align}
\phi^{j_1}(x,\bx,z,\bz)\phi^{j_2}(0,0,0,0)=\sum_{j_3}C^{j_1\,j_2}_{j_3}
(x,\bx,z,\bz)\phi^{j_3}(0,0,0,0)\quad.
\end{align}

It is well-known that all the coefficient functions $C^{j_1\,j_2}_{j_3}
(x,\bx,z,\bz)$ in the expansion (11) can be expressed via the weights(conformal
dimensions) of the primary fields(basic operators) and the structure constants
of Operator Algebra \cite{BPZ,FZ}. The structure constants are defined as
coefficients at the primary fields in the OP expansion\footnote{For simplicity
I set $j\in\R$. However, the generalization to $j\in\C$ is straightforward.}
\begin{align}
\ff{1}(x,\bx,z,\bz)\ff{2}(0,0,0,0)=\sum_{j_3}\frac{|x|^{2(j_1+j_2-j_3)}}
{|z|^{2(\dj{1}+\dj{2}-\dj{3})}}C^{j_1\,j_2}_{j_3}\ff{3}(0,0,0,0)\qquad.
\end{align}

The normalized two and three point functions of the primary fields can be
represented as
\begin{align}
\begin{split}
\langle\fzx{1}\fzx{2}\rangle=&
\delta^{j_1j_2}\frac{|x_{12}|^{4j_1}}{|z_{12}|^{4\dj{1}}}\qquad,\\
\langle\fzx{1}\fzx{2}\fzx{3}\rangle=&C^{j_1\,j_2\,j_3}
\prod_{n<m}\frac{|x_{nm}|^{2\gamma_{nm}(j)}}{|z_{nm}|^{2\gamma_{nm}(\Delta)}}\quad,
\end{split}
\end{align}
where
$y_{nm}=y_n-y_m\,,\,\gamma_{12}(y)=y_1+y_2-y_3\,,\,
\gamma_{13}(y)=y_1+y_3-y_2\,,\, \gamma_{23}(y)=y_2+y_3-y_1\,$.

As to four point function, one can find it in the following form(see \cite{FZ})
\begin{align}
\begin{split}
\langle\fzx{1}\dots\fzx{4}\rangle
=G^{j_1,j_2,j_3,j_4}(x,\bx,z,\bz)
\prod_{n<m}\frac{|x_{nm}|^{2\gamma_{nm}(j)}}{|z_{nm}|^{2\gamma_{nm}
(\Delta)}}\quad,
\end{split}
\end{align}
with
$\gamma_{14}(y)=2y_1,\,\gamma_{23}(y)=y_1+y_2+y_3-y_4,\,
\gamma_{24}(y)=-y_1+y_2-y_3+y_4,\,$
\newline$\gamma_{34}(y)=-y_1-y_2+y_3+y_4$  and
\begin{align*}
x=\frac{x_{12}x_{34}}{x_{14}x_{32}}\,,\qquad\bx=\frac{\bx_{12}\bx_{34}}
{\bx_{14}\bx_{32}}\,,\qquad z=\frac{z_{12}z_{34}}{z_{14}z_{32}}\,,\qquad
\bz=\frac{\bz_{12}\bz_{34}}{\bz_{14}\bz_{32}}\qquad.
\end{align*}

In order to write down (14) explicitly one needs to calculate
$G^{j_1,j_2,j_3,j_4}(x,\bx,z,\bz)$. To this purpose I use the remarkable
relation between the KZ equation \cite{KZ} with the generators of $sl_2$
determined by (6) and the differential equations of degenerate conformal field
theories \cite{BPZ}. Due to this the four point function (14) is expressed
via a five point function of the DF model, where $x,\bx$ play a role of
the coordinate of the fifth operator. For further details I refer to
the original work \cite{FZ}.
Although the relation was discovered by Fateev and Zamolodchikov
for the SU(2) WZW models and the minimal models it proves that the same result
is valid in the case of the degenerate SL(2) conformal field theories and the
Dotsenko-Fateev models($\am^2\in\C$) \cite{B}.

Let $\ff{}(x,\bx,z,\bz)=\ff{}_{+}(x,\bx,z,\bz)$ with
$j=j^+$ and
$\ff{}(x,\bx,z,\bz)=\ff{}_{-}(x,\bx,z,\bz)$ with
$j=j^-$, where $j^{\alpha}$ are defined in (5). The functions
$G^{j_1,j_2,j_3,j_4}(x,\bx,z,\bz)$ are given by
\begin{align}
\begin{split}
G^{(A)}(x,\bx,z,\bz)&=Z^{(A)}(j_1,j_2,j_3,j_4)|z|^a|1-z|^b
\prod_{i=1}^{n_1-1}\prod_{\ip=1}^{m_1-1}\int d^2\,u_i\int d^2\,w_{\ip}\,\,
\vert u_i-w_{\ip}\vert^{-4}\times\\
&\times\prod_{i=1}^{n_1-1}\vert u_i\vert^{4\alpha_2^{(A)}\am}\vert
1-u_i\vert^{4\alpha_3^{(A)}\am} \vert x-u_i\vert^{4\alpha_{21}\am}\vert
z-u_i\vert^{4\alpha_1^{(A)}\am} \prod_{i<\ip}^{n_1-1}\vert
u_{i\ip}\vert^{4\am^2} \times\\
&\times\prod_{i=1}^{m_1-1}\vert
w_{i}\vert^{4\alpha_2^{(A)}\ap}\vert 1-w_i\vert^{4\alpha_3^{(A)}\ap}\vert
x-w_{i}\vert^{4\alpha_{21}\ap} \vert z-w_i\vert^{4\alpha_1^{(A)}\ap}
\prod_{i<\ip}^{m_1-1}\vert w_{i\ip}\vert^{4\ap^2}.
\end{split}
\end{align}
Here $\quad G^{(1)}(x,\bx,z,\bz)=G^{j_1,j_2,j_3,j_4}_{+\,+\,+\,+}(x,\bx,z,\bz),
\quad
G^{(2)}(x,\bx,z,\bz)=G^{j_1,j_2,j_3,j_4}_{+\,+\,+\,-}(x,\bx,z,\bz)$,
\newline
$G^{(3)}(x,\bx,z,\bz)=G^{j_1,j_2,j_3,j_4}_{+\,+\,-\,-}(x,\bx,z,\bz),\,\,
G^{(4)}(x,\bx,z,\bz)=G^{j_1,j_2,j_3,j_4}_{+\,-\,-\,-}(x,\bx,z,\bz),\,\,
a=4j_1j_2\ap^2,$
\newline$b=4j_1j_3\ap^2\,,\,\am=-\sqrt{k+2}\,,\,\ap\am=-1$. Furthermore, the
$\alpha_i^{(A)}$'s are defined via $\alpha_i^{(A)}=\frac{1-N_i^{(A)}}{2}\am+
\frac{1-M_i^{(A)}}{2}\ap$ with

\begin{alignat*}2
N_1^{(1)}&=N_2^{(3)}=
\frac{n_1}{2}+\frac{n_2}{2}+\frac{n_3}{2}+\frac{n_4}{2}-1\,\,
&,\quad
M_1^{(1)}&=M_2^{(3)}=
\frac{m_1}{2}+\frac{m_2}{2}+\frac{m_3}{2}+\frac{m_4}{2}\quad; \\
N_2^{(1)}&=N_1^{(3)}=
\frac{n_1}{2}+\frac{n_2}{2}-\frac{n_3}{2}-\frac{n_4}{2}\,\, &,\quad
M_2^{(1)}&=M_1^{(3)}=
\frac{m_1}{2}+\frac{m_2}{2}-\frac{m_3}{2}-\frac{m_4}{2}\quad;\\
N_3^{(1)}&=-N_4^{(3)}=
\frac{n_1}{2}-\frac{n_2}{2}+\frac{n_3}{2}-\frac{n_4}{2}\,\,
&,\quad
M_3^{(1)}&=-M_4^{(3)}=
\frac{m_1}{2}-\frac{m_2}{2}+\frac{m_3}{2}-\frac{m_4}{2}\,\,; \\
N_4^{(1)}&=-N_3^{(3)}=
-\frac{n_1}{2}+\frac{n_2}{2}+\frac{n_3}{2}-\frac{n_4}{2}\,\,
&,\quad
M_4^{(1)}&=-M_3^{(3)}=
-\frac{m_1}{2}+\frac{m_2}{2}+\frac{m_3}{2}-\frac{m_4}{2};
\end{alignat*}
\begin{alignat*}2
N_1^{(2)}&=-N_4^{(4)}=
\frac{n_1}{2}+\frac{n_2}{2}+\frac{n_3}{2}-\frac{n_4}{2}-\frac{1}{2}\,\,
&,\quad
M_1^{(2)}&=-M_4^{(4)}=
\frac{m_1}{2}+\frac{m_2}{2}+\frac{m_3}{2}-\frac{m_4}{2}\,\,;  \\
N_2^{(2)}&=N_3^{(4)}=
\frac{n_1}{2}+\frac{n_2}{2}-\frac{n_3}{2}+\frac{n_4}{2}-\frac{1}{2}\,\,
&,\quad
M_2^{(2)}&=M_3^{(4)}=
\frac{m_1}{2}+\frac{m_2}{2}-\frac{m_3}{2}+\frac{m_4}{2}\quad;  \\
N_3^{(2)}&=N_2^{(4)}=
\frac{n_1}{2}-\frac{n_2}{2}+\frac{n_3}{2}+\frac{n_4}{2}-\frac{1}{2}\,\,
 &,\quad
M_3^{(2)}&=M_2^{(4)}=
\frac{m_1}{2}-\frac{m_2}{2}+\frac{m_3}{2}+\frac{m_4}{2}\quad;  \\
N_4^{(2)}&=-N_1^{(4)}=
-\frac{n_1}{2}+\frac{n_2}{2}+\frac{n_3}{2}+\frac{n_4}{2}-\frac{1}{2}\,\,
 &,\quad
M_4^{(2)}&=-M_1^{(4)}=
-\frac{m_1}{2}+\frac{m_2}{2}+\frac{m_3}{2}+\frac{m_4}{2}.
\end{alignat*}
In order to take into account a relative normalization between the operators of
the DF models and the ones of the SL(2) degenerate conformal
field theories one has to introduce the normalization constants
$Z^{(A)}(j_1,j_2,j_3,j_4)$. Up to irrelevant factors they may be
written as follows
\begin{align}
\begin{split}
Z^{(1)}(j_1,j_2,j_3,j_4)=V(|N_1^{(1)}|,|M_1^{(1)}|)
\prod_{\{2,3,4\}} V(\vert N_i^{(1)}\vert+1,\vert M_i^{(1)}\vert+1)\quad,
\end{split}
\label{eq:R}\\
\begin{split}
Z^{(2)}(j_1,j_2,j_3,j_4)=V(\vert N_4^{(2)}\vert+1,\vert M_4^{(2)}\vert+1)
\prod_{\{1,2,3\}} V(|N_i^{(2)}|,|M_i^{(2)}|)\quad,
\end{split}
\label{eq:S}\\
\begin{split}
Z^{(3)}(j_1,j_2,j_3,j_4)=V(|N_2^{(3)}|,|M_2^{(3)}|)
\prod_{\{1,3,4\}} V(\vert N_i^{(3)}\vert+1,\vert M_i^{(3)}\vert+1)\quad,
\end{split}
\label{eq:T}\\
\begin{split}
Z^{(4)}(j_1,j_2,j_3,j_4)=V(\vert N_1^{(4)}\vert+1,\vert M_1^{(4)}\vert+1)
\prod_{\{2,3,4\}} V(|N_i^{(4)}|,|M_i^{(4)}|)\quad,
\end{split}
\end{align}
where
\begin{gather*}
V(n,m)=\rho^{-(n-1)(m-1)}P(n,m)\quad,\quad\rho=\ap^2\quad,
\quad\prh=\am^2\quad.
\end{gather*}
\newline The function $P(n,m)$ is given by
\begin{gather*}
P(n,m)=\prod_{i=1}^{n-1}\prod_{j=1}^{m-1}[i\prh-j]^{-2}
\prod_{i=1}^{n-1}\frac{\Gamma[i\prh]}{\Gamma[1-i\prh]}
\prod_{j=1}^{m-1}\frac{\Gamma[j\rho]}{\Gamma[1-j\rho]}\quad ,\quad
P(1,1)=1\quad.
\end{gather*}

The 2D multiple integrals (15) are defined via contour integrals. This can be
done by using the results of Dotsenko and Fateev \cite{DF,D1}.

The structure constants are found in the form
\begin{align}
\begin{split}
C^{j_1,j_2,j_3}_{+\,+\,+}&=\frac{Z^{(1)}(0,j_1,j_2,j_3)
\biggl( Z^{(1)}(0,0,0,0)\biggr)^{\h}}
{\biggl( Z^{(1)}(0,j_1,j_1,0)Z^{(1)}(0,j_3,0,j_3)
Z^{(1)}(0,0,j_2,j_2)\biggr)^{\h}}=\\
&=\biggl(\frac{\Gamma[\rho]}{\Gamma[1-\rho]}\biggr)^{\h}
P(\sp-\h,\sigma+\h)\times \\
&\times\prod_{\{1,2,3\}}(-)^{\frac{n_i-1}{2}}\rho^{(1-n_i)}
\biggl(\frac{\Gamma[n_i-m_i\rho]}{\Gamma[1-n_i+m_i\rho]}\biggr)^{\h}
\frac{P(\sp-n_i+\h,
\sigma-m_i+\h)}{P(n_i,m_i)}\quad,
\end{split}
\end{align}
\begin{align}
\begin{split}
C^{j_1,j_2,j_3}_{+\,+\,-}&\propto Z^{(2)}(0,j_1,j_2,j_3)=\\
&=\rho^{-\h}P(\sp,\sigma+\h)\prod_{\{1,2,3\}}
\rho^{-(n_i-1)(m_i-\h)}P(\sp-n_i,\sigma-m_i+\h)\quad,
\end{split}
\label{eq:K}\\
\begin{split}
C^{j_1,j_2,j_3}_{+\,-\,-}&\propto \frac{Z^{(3)}(0,j_1,j_2,j_3)}
{\biggl( Z^{(3)}(0,0,j_2,j_2)Z^{(3)}(0,0,j_3,j_3)\biggr)^{\h}}=\\
&=\rho^{-(n_1-1)(m_1-\h)}\biggl(P(n_1,m_1+1)P(n_1,m_1)\biggr)^{\h}
P(\sp-\h,\sigma+\h)\times\\
&\times\prod_{\{1,2,3\}}\frac{P(\sp-n_i+\h,\sigma-m_i+\h)}
{\biggl(P(n_i,m_i+1)P(n_i,m_i)\biggr)^{\h}}\quad,
\end{split}
\label{eq:L}\\
\begin{split}
C^{j_1,j_2,j_3}_{-\,-\,-}&\propto Z^{(4)}(0,j_1,j_2,j_3)=\\
&=\rho^{-\h}P(\sp,\sigma+\h)\prod_{\{1,2,3\}}
\rho^{-(n_i-1)(m_i-\h)}P(\sp-n_i,\sigma-m_i+\h)\quad.
\end{split}
\end{align}
Here $\sigma^{\prime}=\frac{n_1}{2}+\frac{n_2}{2}+\frac{n_3}{2}$ and $\sigma=
\frac{m_1}{2}+\frac{m_2}{2}+\frac{m_3}{2}$. It should be noted that the
normalization of (21-23), as defined in (13), is ambiguous because $\Phi_-$
operators don't contain the unity operator. It belongs to the $\Phi_+$
operators, namely $\bf 1=\Phi_{+}^{j=0}$.

Setting $x=z,\bx=\bz$ in (15), which corresponds to the quantum hamiltonian
reduction \cite{PG}, one immediately obtain the relation between the proper
4-point functions of the DF model(see (28)). Due to $Z_2$ symmetry of the model
it is sufficient to consider the $\Phi_+$ fields only. Using (20) I get the
structure constants in the following symmetric form
\begin{align}
\begin{split}
C(n_1,m_1;n_2,m_2;n_3,m_3)&=P^{\h}(2,2)\,P(\sp+\h,
\sigma+\h)\times\\
&\times\prod_{\{1,2,3\}}\frac{P(\sp-n_i+\h,
\sigma-m_i+\h)}{\biggl(P(n_i+1,m_i+1)P(n_i,m_i)\biggr)^{\h}}\quad.
\end{split}
\end{align}

{}From the set (5) it is worth to distinguish the so-called admissible
representations \cite{KW}, which correspond to the rational level k. In the
case $k=-2+p/q$, with the coprime integers $p$ and $q$, it is possible to
recover the minimal models (series with $c<1$ ) via the Drinfeld-Sokolov
reduction. On the other hand $k=-2-p/q$ leads to the Liouville series with
$c>25$. The second point is an existence of modular invariants for such
representations.

At the rational level $k=-2+p/q$ there is a symmetry
$j^-_{n,m}=j^+_{q-n+1,p-m}$ which allows one to reduce the $\Phi_-$ fields
to the $\Phi_+$ ones.
Up to the normalization factors one can identify the various C's, namely
$C^{j_1,j_2,j_3}_{+\,+\,+}=C^{j_1,j_2,j_3}_{+\,-\,-}$ and
$C^{j_1,j_2,j_3}_{+\,+\,-}=C^{j_1,j_2,j_3}_{-\,-\,-}$.
This results in the following structure constants
\begin{equation}
\begin{split}
C^{j_1,j_2,j_3}&=C^{j_1,j_2,j_3}_{+\,+\,+}\quad, \\
\text{and}&\\
C^{j_1,j_2,j_3}&\propto C^{j_1,j_2,j_3}_{-\,-\,-}\quad\text{with}\quad
n_i\rightarrow q-n_i+1,\,\,m_i\rightarrow p-m_i\,.
\end{split}
\end{equation}
It is easy to see from (25) that the OP algebra at the rational
level is closed in the grid $1\leq n_i\leq q,\,\,1\leq m_i\leq p-1$.  The
corresponding fusion rules are given by \begin{align} \begin{cases}
\text{max}\,(n_{12}+1,\,n_{21}+1)\,\leq\, n_3\leq\text{min}\,(n_1+n_2-1,\,
2q-n_1-n_2+1)\quad,\\
\text{max}\,(m_{12}+1,\,m_{21}+1)\leq m_3\leq\text{min}\,(m_1+m_2-1,\,
2p-m_1-m_2-1)\quad,
\end{cases}\\
\vspace{0.5cm}
\begin{cases}
\text{max}\,(n_1+n_2-q+1,\,q+3-n_1-n_2)\,\leq\, n_3\leq\text{min}\,
(n_{12}+q-1,\,n_{21}+q-1)\quad,\\
\text{max}\,(m_1+m_2-p+1,\,p+1-m_1-m_2)\leq m_3\leq\text{min}\,
(m_{12}+p-1,\,m_{21}+p-1)\quad,
\end{cases}
\end{align}
As a checking procedure one can directly analyze singularities of (15).
These fusion rules agree with those found in \cite{AY} from the differential
equations for the conformal blocks. Later the same result was obtained by
cohomological methods \cite{FM}.
\newline It should be also noted that in the case of
$n_i=1,\,\rho=\frac{1}{k+2},\,k\in\N\,\,\text{and} \,\,m_i=2j_i+1$
the structure constants and fusion rules of the unitary representations are
recovered \cite{FZ}.

Let me now further investigate the theory at the rational level. In order to
consider the level $k$ defined by $k=-2-p/q$ one can proceed in complete
accordance with the previous case. A simple analysis shows the same fusion
rules as (26-27). As to the structure constants, they look like (25). The only
difference is the sign of $\rho(\prh)$.

Let me now conclude by giving some remarks.
\newline
(i) One is on the normalizations of
the structure constants (21-23). Note that the usual normalization (13) is
useless for this purpose, therefore it is necessary to look for something more.
One can try to solve the bootstrap equations \cite{BPZ}.
It could clarify the problem of picking up the right normalization, but this is
beyond the scope of this note.
\newline (ii) Another remark is on the Wakimoto free field
representation for these models. The first attempts were made to extend the
Wakimoto representation to the rational weights in \cite{BF,D}. However all of
them used the ordinary construction attaching representation to a point, i.e.
without $x$. As a result their OP expansions were ambiguous, so the fusion
rules are determined up to $j=-j-1$ identification. Recently, Petersen,
Rasmussen and Yu developed the Wakimoto representation that relies upon
introducing the isotopic coordinates related to the $sl_2$ representations
\cite{PRY}. Using that construction, they found the conformal blocks on the
sphere and the fusion rules like (26-27). The next step would be to build the
correlation functions and to derive the OP algebra of the primary fields.  The
theory is then fully solved by the Wakimoto representation. Unfortunately the
last step have not realized yet.  \newline (iii) Impressive seems the following
relation between the correlations functions of the DF model
\begin{align}
\langle\mo{n_1}{m_1}\mo{n_2}{m_2}\mo{n_3}{m_3}\mo{n_4}{m_4}\rangle
\,\propto\,\langle\mo{N_1}{M_1}\mo{N_2}{M_2}\mo{N_3}{M_3}\mo{N_4}{M_4}\rangle\quad,
\end{align}
where
\begin{alignat*}2
N_1^{(1)}&=
\frac{n_1}{2}+\frac{n_2}{2}+\frac{n_3}{2}+\frac{n_4}{2}\,\,
&,\qquad
M_1^{(1)}&=
\frac{m_1}{2}+\frac{m_2}{2}+\frac{m_3}{2}+\frac{m_4}{2}\quad; \\
N_2^{(1)}&=
\frac{n_1}{2}+\frac{n_2}{2}-\frac{n_3}{2}-\frac{n_4}{2}\,\, &,\qquad
M_2^{(1)}&=
\frac{m_1}{2}+\frac{m_2}{2}-\frac{m_3}{2}-\frac{m_4}{2}\quad;
\\
N_3^{(1)}&=
\frac{n_1}{2}-\frac{n_2}{2}+\frac{n_3}{2}-\frac{n_4}{2}\,\,
&,\qquad
M_3^{(1)}&=
\frac{m_1}{2}-\frac{m_2}{2}+\frac{m_3}{2}-\frac{m_4}{2}\quad; \\
N_4^{(1)}&=
-\frac{n_1}{2}+\frac{n_2}{2}+\frac{n_3}{2}-\frac{n_4}{2}\,\,
&,\qquad
M_4^{(1)}&=
-\frac{m_1}{2}+\frac{m_2}{2}+\frac{m_3}{2}-\frac{m_4}{2}\,\,.
\end{alignat*}
The problem is to understand what underlies this mysterious relation. May be
there is a hidden symmetry in the theory.

I am grateful to Vl.Dotsenko and B.Feigin for fruitful discussions and
A.Semikhatov for reading the manuscript. I would also like to thank
M.Lashkevich and V.B.Petkova for comments. This work was supported in part by
Russian Basic Research Foundation under grant 93-02-3135.


\begin{thebibliography}{99}

\bibitem{BPZ}
A.A.Belavin, A.M.Polyakov and A.B.Zamolodchikov, \NP{241}{1984} 333.

\bibitem{DF}
Vl.S.Dotsenko and V.A.Fateev, {\it Nucl.Phys.}{\bf B240 [FS12]} (1984) 312;
\newline{\it Nucl.Phys.}{\bf B251 [FS13]} (1985) 691; \PL{157}{1985} 291.

\bibitem{FZ}
V.A.Fateev and A.B.Zamolodchikov, {\it Sov.J.Nucl.Phys.}{\bf 43} (1986) 657.

\bibitem{CF}
P.Christe and R.Flume, \PL{188}{1987} 219;
V.B.Petkova and J.-B.Zuber, \NP{438}{1995} 347.

\bibitem{KK}
V.G.Kac and D.A.Kazhdan, {\it Adv.Math.} {\bf 34} (1979) 97.

\bibitem{FM}
B.Feigin and F.Malikov, Fusion algebra at a rational level and cohomology
of nilpotent subalgebras of $\ls$, hep-th/9310003.

\bibitem{PG}
P.Furlan, A.Ch.Ganchev, R.Panov and V.B.Petkova, \PL{267}{1991} 63;
\NP{394}{1993} 665.

\bibitem{KZ}
V.G.Knizhnik and A.B.Zamolodchikov, \NP{247}{1984} 83.

\bibitem{B}
B.Feigin, Private communication.

\bibitem{D1}
Vl.S.Dotsenko, {\it Adv.Stud. in Pure Math.} {\bf 16} (1988) 123.

\bibitem{KW}
V.G.Kac and M.Wakimoto, {\it Proc.Natl.Acad.Sci.USA} {\bf 85} (1988) 4956.

\bibitem{AY}
H.Awata and Y.Yamada, {\it Mod.Phys.Lett.} {\bf A7} (1992) 1185.

\bibitem{BF}
D.Bernard and G.Felder, \CMP{127}{1990} 145.

\bibitem{D}
Vl.S.Dotsenko, \NP{338}{1990} 747; \NP{358}{1991} 541.

\bibitem{PRY}
J.L.Petersen, J.Rasmussen and M.Yu, Conformal blocks for admissible
representations in $SL(2)$ current algebra, Preprint NBI-HE-95-16,
hep-th/9504127.

\end{thebibliography}
\end{document}